# Evidence of Mott Insulator with Thermally Induced Melting Behavior in Kagome Compound $Nb_3Cl_8$

Qiu Yang[1†], Min Wu[1†*], Jingyi Duan[2,3†], Zhijie Ma[4], Lingxiao Li[1], Zihao Huo[1], Zaizhe Zhang[1], Kenji Watanabe[5], Takashi Taniguchi[6], Xiaoxu Zhao[7], Yi Chen[1], Youguo Shi[4], Wei Jiang[2,3], Kaihui Liu[8*], Xiaobo Lu[1,9*]

[1]International Center for Quantum Materials, School of Physics, Peking University, Beijing 100871, China
[2]Centre for Quantum Physics, Key Laboratory of Advanced Optoelectronic Quantum Architecture and Measurement (MOE), School of Physics, Beijing Institute of Technology, Beijing 100081, China
[3]Beijing Key Lab of Nanophotonics & Ultrafine Optoelectronic Systems, School of Physics, Beijing Institute of Technology, Beijing, 100081, China
[4]Beijing National Laboratory for Condensed Matter Physics, Institute of Physics, Chinese Academy of Sciences, Beijing, China
[5]Research Center for Electronic and Optical Materials, National Institute of Material Sciences, 1-1 Namiki, Tsukuba 305-0044, Japan
[6]Research Center for Materials Nanoarchitectonics, National Institute of Material Sciences, 1-1 Namiki, Tsukuba 305-0044, Japan
[7]School of Materials Science and Engineering, Peking University, Beijing, 100871, China
[8]State Key Laboratory for Mesoscopic Physics, Frontiers Science Centre for Nano-optoelectronics, School of Physics, Peking University, Beijing 100871, China
[9]Collaborative Innovation Center of Quantum Matter, Beijing 100871, China

[†]Those authors contribute equally to this work.
[*]Corresponding Authors: min.wu@pku.edu.cn, khliu@pku.edu.cn, xiaobolu@pku.edu.cn

## ABSTRACT

The kagome lattice provides a playground to explore novel correlated quantum states due to the presence of flat bands in its electronic structure. Recently discovered layered kagome compound $Nb_3Cl_8$ has been proposed as a Mott insulator coming from the half-filled flat band. Here we have carried out systematic transport study to uncover the evidence of Mott insulator in $Nb_3Cl_8$ thin flakes. Bipolar semiconducting property with Fermi level close to conduction band has been revealed. We have further probed the chemical potential of $Nb_3Cl_8$ by tracing the charge neutrality point of monolayer graphene proximate to $Nb_3Cl_8$. The gap of $Nb_3Cl_8$ flakes is approximately 1.10 eV at 100 K and shows pronounced temperature dependence, decreasing substantially with increasing temperature to ~0.63 eV at 300 K. The melting behavior of the gapped state is in consistent with theoretically proposed Mott insulator in $Nb_3Cl_8$. Our work has demonstrated $Nb_3Cl_8$ as a promising platform to study strongly correlated physics at relatively high temperature.

## INTRODUCTION

Electron correlation-induced Mott insulators serve as parent phases for numerous strongly correlated phenomena, including high temperature superconductor and magnetism [1,2]. Mott insulators are commonly observed in materials such as layered perovskites (e.g., $La_2CuO_4$ [3–5], $Sr_2IrO_4$ [6–8]), monometallic oxides (e.g., $V_2O_3$ [9,10], NiO [11–13]), transition metal dichalcogenides (e.g., $NiS_2$ [14,15], 1T-$TaS_2$ [16–18]) and so on. These systems exhibit complex many-body ground states that cannot be explained by conventional band theory of solids. A hallmark feature of Mott insulators is the half-filled electronic band structure near the Fermi level, where the strong on-site Coulomb interactions localize the electrons, preventing charge transport and opening a Mott gap [19–21]. However, achieving precise control over the electronic state in conventional Mott insulators often relies on extreme conditions (such as high pressure, cryogenic temperatures) or irreversible methods such as chemical doping, thereby limiting potential applications of Mott materials.

The system exhibiting flat band offer a promising route toward realizing Mott insulating states, where Coulomb interactions significantly stronger than kinetic energy, readily enabling strong electron correlations, such as in artificial superlattices [22–24]. Alternatively, flat band structures can be naturally engineered in one promising route, the kagome lattice [25–27]. Kagome lattice is a geometrically frustrated structure constituted by corner sharing triangles, with the coexistence of Dirac cone, van Hove singularity, and flat bands in its electronic structure, making this system a fertile platform to investigate the various quantum phenomena originating from the interplay between topology, geometry and correlation [28–30]. However, the correlated phenomena associated with the flat bands remain largely underexplored. One prime reason is that the flat bands in the metallic kagome materials lie away from the Fermi surface and intertwine with other bands [25,31–34], which precludes the exploration of correlated behavior contributed from the flat bands. Additionally, the difficulty in producing ultrathin flakes [35] (although the $AV_3Sb_5$ family is a van der Waals material [36]) and the absence of band gap in kagome metals both limit their potential applications in nano-electronic devices. Thus, layered semiconducting kagome materials with isolated flat bands near the Fermi level are exceptionally interesting.

Recently a new family of van der Waals kagome compounds, $Nb_3X_8$ ($X$= Cl, Br, I), has been proposed as an ideal system to study the strongly correlated physics [37–43]. In these layered materials, the Nb atoms in each layer form trigonally distorted kagome lattice, namely breathing kagome lattice, as shown in Figs. 1(a) and 1(b) for $Nb_3Cl_8$. An Angle-resolved photoemission spectroscopy (ARPES) experiments with calculations of $Nb_3Cl_8$ crystals have revealed that the flat band separated from other bands, away from the Fermi level, and a single-particle band gap arising from symmetry breaking opened at the Fermi surface [38]. However, the latest ARPES measurements and calculations present different results, the flat band lies near the Fermi level and is half-filled, giving rise a Mott gap that can be described by single-band Hubbard model [40]. Understanding the nature of the ground state is very instructive for further exploring the fascinating electronic phases in $Nb_3Cl_8$.

In this work, we fabricated hexagonal boron nitride (hBN)-encapsulated dual-gated devices of $Nb_3Cl_8$ thin flakes. This allows us to directly measure the resistance modulated by the gate voltages and assess its relationship with temperature. Electronic transport measurements reveal that $Nb_3Cl_8$ exhibits semiconducting behavior with ambipolar characteristics. To investigate the nature of band gap, monolayer graphene (MLG) was employed to monitor the chemical potential $\mu$ of $Nb_3Cl_8$. A gap of $\sim 1.10\ \mathrm{eV}$ was observed at $100\ \mathrm{K}$ via tracking the charge neutrality point (CNP) of MLG, exhibiting significant temperature dependence with rapid reduction at elevated temperatures. This behavior strongly contrasts with conventional single-particle gap, in which the gap remains nearly temperature-independent, and can be attribute to the emergence of a Mott gap driven by strong electron correlations.

## RESULTS

Figure 1(c) shows the non-interacting band structure of bulk $Nb_3Cl_8$ with a bilayer stacking periodicity, where flat bands cross the Fermi level in individual layers with half-filled occupation states. Notably, the number of half-filled flat bands at the Fermi surface in the single-particle framework equals to the stacking periodicity in bulk $Nb_3Cl_8$. Due to the exceptionally weak coupling between neighboring layers, the electronic properties of each layer are primarily dominated by intralayer characteristics. The flat band in monolayer $Nb_3Cl_8$ is also half-filled across the Fermi level without electron correlations (as shown in Fig. S4). $Nb_3Cl_8$ thin flakes can be readily mechanically exfoliated from the bulk crystal, Figure 1(d) shows the optical image of a typical $Nb_3Cl_8$ thin flake with thickness about $2.6\ \mathrm{nm}$ determined by atomic force microscope (AFM) (Fig. 1(e)). The hBN-encapsulated dual-gated device, as schematically illustrated in Fig. 1(f), were fabricated using dry transfer techniques and standard nanofabrication techniques. The two-terminal conductance of a typical $2.6\ \mathrm{nm}$-thick sample S1 (inset in Fig. 1(g)), as a function of gate voltages, exhibits a bipolar semiconducting characteristic with slightly electron-doped, as shown in Fig. 1(g). Similar behavior has been observed in device S2, which has a thickness of $4.2\ \mathrm{nm}$ (see Fig. S3).

Interestingly, even when being highly doped with electrons, i.e. at $V_{\mathrm{tg}} = V_{\mathrm{bg}} = 10\ \mathrm{V}$, corresponding to a carrier density about $n = 1.3 \times 10^{13}\ \mathrm{cm}^{-2}$, the sample exhibits robust insulating behavior with clear thermal activation observed over a wide temperature range $(100 -$ 260 K). Using the Arrhenius formula $R \propto \exp[-\Delta/2k_B T]$, where $k_B$ is the Boltzmann constant, the thermal activation gap $\Delta$ as a function of $V_{tg}$ is quantitatively depicted in Fig. 1(h) with the insert shows the two-terminal resistance versus temperature at different $V_{tg}$. We note that the four-terminal measurements (shown in Fig. S2) illustrate essentially the same features, including the magnitude and variation of $\Delta$ with $V_{tg}$ as marked by the pink dots in Fig. 1(h). Therefore, we consider the two-terminal data presented here to be compelling and provides a definitive representation of the fundamental properties of $Nb_3Cl_8$ flakes.

Surprisingly, a gap up to $\sim 105$ meV can still be revealed at $V_{tg} = V_{bg} = 10$ V, highly indicating the presence of strong electron interactions. The interaction strength can be qualitatively described by ratio between potential energy and kinetic energy, which can be expressed as $r_s = \frac{n_v m^* e^2}{4\pi\varepsilon h^2 \sqrt{\pi n}}$, where $n_v$ is the number of degenerate valleys, $\varepsilon$ is the dielectric constant and $m^*$ is the effective electron mass [44]. For $Nb_3Cl_8$, we calculated $r_s \approx 24.8$ (60.9) for valence (conduction) band , with $n_v = 1$, $m^* = -2.53m_0$ ($6.20m_0$), $\varepsilon = 3\varepsilon_0$(the effective electron mass and dielectric constant are calculated from the band structure of $Nb_3Cl_8$ and the geometric capacitance of the MLG/hBN/$Nb_3Cl_8$ system, respectively, as described in Method), and $n = 1.3 \times 10^{13}\ \mathrm{cm}^{-2}$ for the largest carrier density in our experiments. The ultra-high $r_s$ value confirms $Nb_3Cl_8$ is a strongly correlated system, potentially giving rise to exotic phenomena such as the existence of an intrinsic Wigner crystal.

Furthermore, to investigate the nature of the gap, monolayer graphene (MLG) was employed to probe the chemical potential $\mu$ of $Nb_3Cl_8$, as shown in the schematic diagram of the measurement configuration in Fig. 2(a). Such a methodology has been widely used to track the chemical potential in various two-dimensional systems [45–48]. In the heterostructure devices, MLG and $Nb_3Cl_8$ thin flake are separated by a thin hBN layer (~ 5 nm). Fig. 2(b) shows the band alignment of MLG and $Nb_3Cl_8$ with the control of double gates. Here, we set $\mu_{MLG} = 0$ when the carrier density of MLG, $n_{MLG} = 0$, corresponding to the charge neutrality point (CNP). Then the chemical potential $\mu_{Nb_3Cl_8}$ and carrier density $n_{Nb_3Cl_8}$ of $Nb_3Cl_8$ flake are given by

$$\mu_{Nb_3Cl_8} = -\frac{eC_{tg}V_{tg}}{C_{eff}} \tag{1}$$

$$n_{Nb_3Cl_8} = \frac{C_{bg}V_{bg}}{e} + \frac{\left(C_{bg} + C_{eff}\right)C_{tg}V_{tg}}{eC_{eff}} \tag{2}$$

where $C_{eff}$ is the effective geometric capacitances per unit area considering both the dielectric properties of hBN and $Nb_3Cl_8$, $e$ is the elementary charge, $C_{tg}$ and $C_{bg}$ are the geometric capacitances per unit area of top hBN and bottom hBN, respectively (See Methods for detailed derivation). Apparently, the evolution of CNP in MLG can reflect carrier density $n_{Nb_3Cl_8}$- dependent chemical potential $\mu_{Nb_3Cl_8}$ of $Nb_3Cl_8$ flake.

Figure 2(c) presents four-terminal resistance of MLG as a function of $V_{tg}$ and $V_{bg}$ at $T = 200$ K. Based on carrier doping, the 2D color map can be divided into nine regions, as shown in Fig. 2(c), and the corresponding band alignments between MLG and $Nb_3Cl_8$ are depicted in Fig. 2(d). The doping states of $Nb_3Cl_8$ are divided into three distinct regions: hole-doped, charge-neutral, and electron-doped (from left to right), based on the two white dashed lines at $V_{bg} = -2.11$ V and $-0.35$ V. The resistance at the Dirac point of MLG varies depending on the type of carrier doped into $Nb_3Cl_8$, as shown on the right side of Fig. 2(c). Owing to the screening effect from MLG, the Fermi level of $Nb_3Cl_8$ is slightly dependent on the top gate $V_{tg}$. For MLG, the CNP is characterized by resistance peak, which clearly defines the transition between electron and hole doping regions. Noted that, the track of CNP is not pass through the point at $V_{tg} = 0$ and $V_{bg} = 0$, which can be attributed to the type-III band alignment between MLG and $Nb_3Cl_8$, as illustrated in Fig. 2(d)-IX. When MLG was placed closed to $Nb_3Cl_8$, electrons in MLG will spontaneously transfer into $Nb_3Cl_8$ even though these two thin flakes were isolated by the spacer layer hBN, similar to the graphene/hBN/$RuCl_3$ device previously reported [49]. In this case, the Fermi level for MLG should be positioned in valence band instead of CNP at initial state ($V_{tg} = 0,\ V_{bg} = 0$) due to charge transfer behavior.

The phase diagram of MLG resistance versus both gate voltage (Fig. 2(c)) can be transformed into parameter space of $V_{tg}$-$n_{Nb_3Cl_8}$ using equations (1) ad (2), as plotted in Fig. 3(a), which conveniently illustrates the evolution of the chemical potential $\mu_{Nb_3Cl_8}$ with continuous hole or electron doping at $T = 200$ K. The chemical potential of $Nb_3Cl_8$ ($\mu_{Nb_3Cl_8}$) can be determined by tracking the CNP position of MLG, and $\mu_{Nb_3Cl_8}$ exhibits a linear dependence on $V_{tg}$ applied to the CNP region, as theoretically illustrated by equations (1). Figure 3(b) shows the variation of the chemical potential $\mu_{Nb_3Cl_8}$ as a function of $n_{Nb_3Cl_8}$ by tracking CNP of MLG (solid white line in Fig. 3(a)). On the electron side, $\mu_{Nb_3Cl_8}$ saturates at approximately about $-0.54$ eV. As carrier density decreases, $\mu_{Nb_3Cl_8}$ sharply decreases and approaches $-1.40$ eV on the hole side. Whereupon, the magnitude of gap $\Delta\mu$ in $Nb_3Cl_8$ is about $0.86$ eV at $200$ K, which is consistent with the theoretical calculations and the optical experiments [38,40]. The $\mu_{Nb_3Cl_8}$ is almost a constant (Figs. 3 and 4(a)) whether on the electron or hole side, which implies the existence of flat bands near the Fermi level in $Nb_3Cl_8$. It is noted that the jump in $\mu_{Nb_3Cl_8}$ occurs around $n_{Nb_3Cl_8} = 1 \times 10^{12} \mathrm{cm}^{-2}$, rather than $n_{Nb_3Cl_8} = 0$, and the extracted $\mu_{Nb_3Cl_8}$ is always negative, which further demonstrates the charge transfer doping in MLG/hBN/$Nb_3Cl_8$ heterostructure.

Figure 4(a) displays the temperature-dependent phase diagram of MLG measured from $100$ K to $300$ K, with the corresponding $\Delta\mu_{Nb_3Cl_8}$ plotted in Fig. 4(c), marked by red stars. Obviously, the magnitude of gap $\Delta\mu_{Nb_3Cl_8}$ is about $1.10$ eV at $100$ K, and then reduced to $0.63$ eV at $300$ K. Similar results were also observed in device M2, as marked with red dots. This dramatic decreasing phenomenon with increasing temperature is strong contrast to conventional semiconductors, where the band gaps change only slightly with temperature increasing [50,51]. To further confirm the difference between $Nb_3Cl_8$ and conventional semiconductor, we probe the chemical potential $\mu$ of bilayer $WSe_2$, whose band gap is about 1.2 eV [52,53]. Similar to the structure of MLG/hBN/$Nb_3Cl_8$ heterostructure device, bilayer $WSe_2$ and MLG are separated by a thin hBN layer. Fig.4(b) shows the four-terminal resistance of MLG on parameter space of $V_{\mathrm{tg}}$-$n_{\mathrm{WSe_2}}$ from 100 K to 300 K on device MLG/hBN/$WSe_2$ heterostructure. Clearly, the CNP of MLG crosses $V_{\mathrm{tg}} = 0$ and $n_{\mathrm{WSe_2}} = 0$, in contrast to that observed in Figs. 2(c) and 3(a), which is accordance with the band alignment between MLG and $WSe_2$ [54]. Similarly, the chemical potential of $WSe_2$ can be obtained by the CNP of MLG. As shown in Fig. 4(c), as marked with blue triangles, the band gap in bilayer $WSe_2$ is about 1 eV and nearly unchanged with temperature.

The evolution of band gap in $Nb_3Cl_8$ with temperature is reminiscent of strong electron interactions generated Mott insulator, in which the rate of band gap reduction is significantly faster than that in conventional semiconductors. For the latter, the band gap is primarily governed by the static electronic structure, with minimal temperature dependence. However, the Mott gap is highly temperature-dependent due to the strong electron correlations. At elevated temperatures, thermal fluctuations may disrupt the strong electron correlations, thereby facilitating the previously suppressed electronic transitions and leading to a significant reduction in the Mott gap, with the decrease in gap size being an order of magnitude larger than $k_{\mathrm{B}}T$ [55,56].

Figures 4(d) and 4(e) show the band structures for both monolayer and bulk $Nb_3Cl_8$ based on density functional theory plus Hubbard U correction (DFT+U), respectively. After considering the electron correlations, the half-filled flat band near the Fermi level splits into upper and lower Hubbard bands, giving rise to a Mott gap. Similar to the band structure in Fig.1(c), the number of upper and lower Hubbard bands in the single-particle framework equals to the stacking periodicity in bulk $Nb_3Cl_8$. When the chemical potential of $Nb_3Cl_8$ is pinned at $-1.40$ eV in Fig.3(b), the Fermi level remains at the lower Hubbard band, as shown in Fig.3(e). As the carrier density increases, the Fermi level crosses to the gap and eventually reaches the upper Hubbard band, as shown in Fig.3(d) and Fig.3(c). Noted that, the Figs. 3(c)-3(e) serve as effective schematics to illustrate the evolution of the electronic structure in the low-doping regime. When the doping level becomes higher, the Fermi level will not simply enter the lower or upper Hubbard bands, instead the electronic structure will reshuffle. Whereupon, the observed insulating behavior at high temperatures, significant temperature-dependent band gap reduction, and band structure calculations collectively provide compelling evidence of Mott insulating behavior in $Nb_3Cl_8$, which persists up to room-temperature. Furthermore, the Mott state is related to the localization of one electron on each $Nb_3$ cluster yields a $[Nb_3]^{8+}$ valence state with localized S = 1/2 moments [57]. This cluster magnetism makes $Nb_3Cl_8$ a promising platform for exploring exotic magnetic phases, such as quantum spin liquids.

## DISCUSSION

In summary, we performed systematic electronic transport measurements to reveal the evidence of Mott states in $Nb_3Cl_8$. The bipolar semiconducting characteristic was observed by direct measurements on $Nb_3Cl_8$ flakes. By employing MLG as the detector layer, the gap size of $Nb_3Cl_8$ flakes, extracted from the chemical potential difference between the hole and electron doped regions, is highly sensitive to temperature. This behavior strongly contrasts with conventional semiconductors, and is attributed to the formation of the Mott gap. The room-temperature Mott insulating behavior in $Nb_3Cl_8$ provides a promising platform for investigating strongly correlated physics as well as moiré engineering in the future.

## ACKNOWLEDGMENTS

This work was supported by the National Key R&D Program (Grant Nos. 2022YFA1403500, 2024YFA1409002 and 2024YFA140840), the National Natural Science Foundation of China (Grant Nos. 12274006, 12141401 and 12404044), Guangdong Major Project of Basic and Applied Basic Research (2021B0301030002), National Key Research and Development Program of China (No. 2024YFA140840), Synergetic Extreme Condition User Facility (SECUF) and Postdoctoral Science Foundation Grant (No. 2023M730100).

*Chemistry* 2017; **56**: 3483–8.

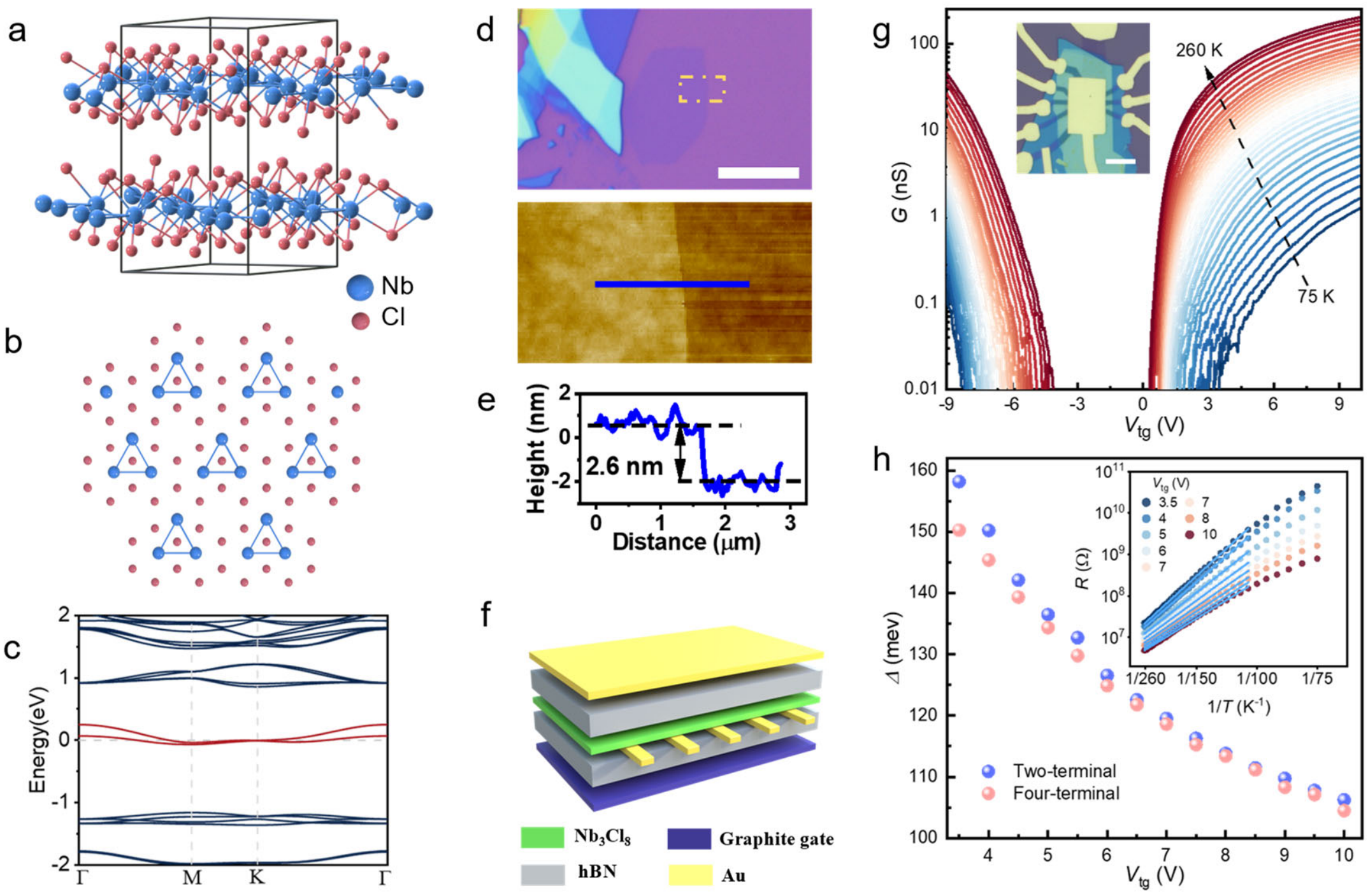

**Figure 1. Fundament properties of $Nb_3Cl_8$.** (a, b) Crystal structure of $Nb_3Cl_8$. The Nb atoms in each layer form trigonally distorted kagome lattice. (c) Non-interacting band structure of bulk $Nb_3Cl_8$ with a bilayer stacking periodicity. (d, e) An optical image of the exfoliated thin flake of $Nb_3Cl_8$ and the corresponding thickness is about 2.6 nm measured by AFM. Scale bar is 10 μm. (f) Schematic of hBN-encapsulated dual-gated device of $Nb_3Cl_8$. The top gate and electrodes are made of Au. (g) The two-terminal conductance of $Nb_3Cl_8$ as a function of top gate voltage. Insert: Optical image of the device S1, scale bar is 10 μm. (h) The thermal activation gap measured with two-terminal (blue dots) and four-terminal (pink dots) configuration as a function of top gate voltage. Insert shows the experimental data from 260 K to 100 K fitted by Arrhenius formula. During the measurements of device S1, the applied bottom gate voltage is equal to the top gate voltage.

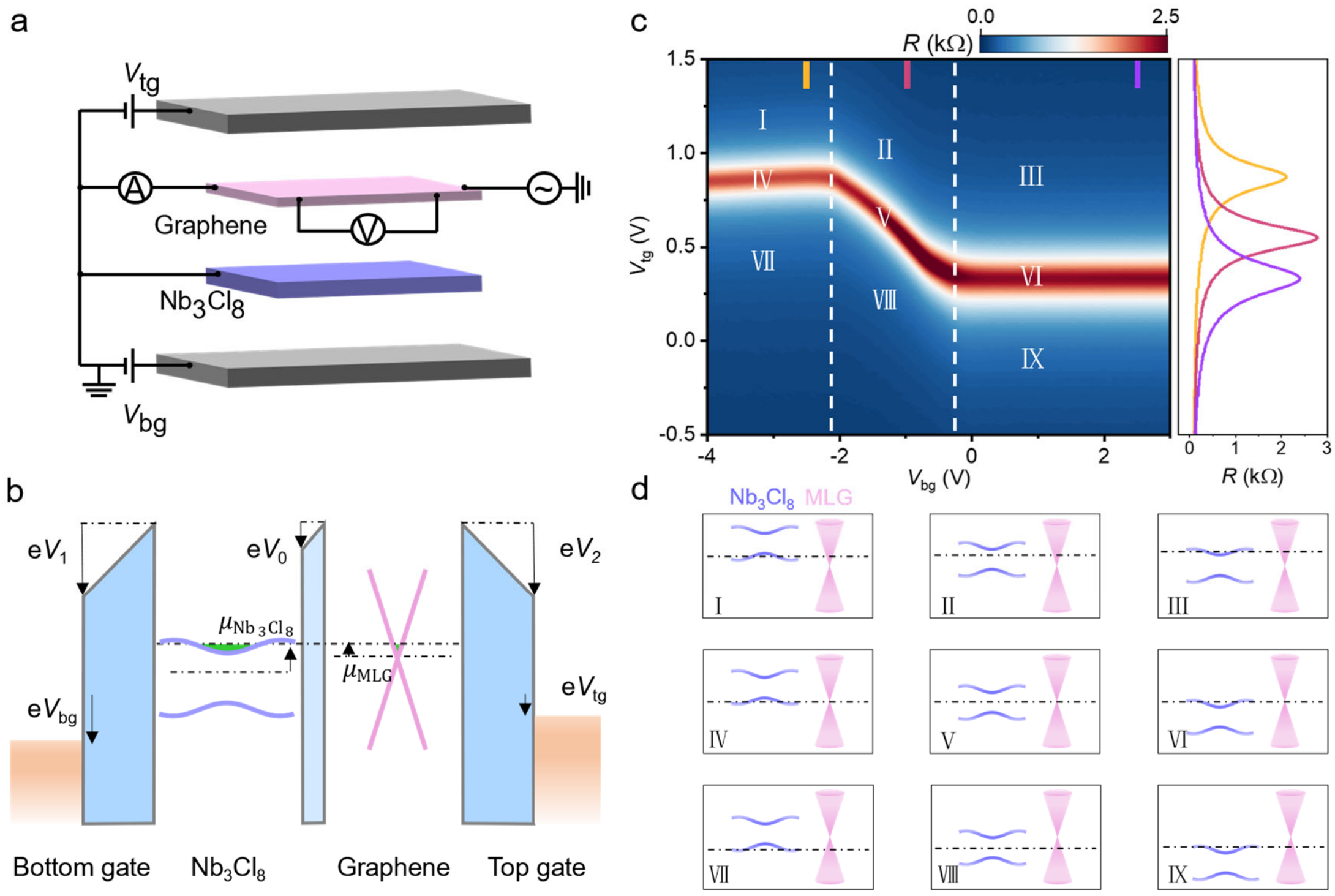

**Figure 2. MLG/hBN/$Nb_3Cl_8$ heterostructure device.** (a) Schematic of the methodology. The MLG and $Nb_3Cl_8$ are separated by a thin hBN layer. (b) Band alignment of MLG and $Nb_3Cl_8$, illustrating the relationship between the chemical potentials of MLG ($\mu_{\mathrm{MLG}}$) and $Nb_3Cl_8$ ($\mu_{\mathrm{Nb_3Cl_8}}$), with the control of top gate voltage ($V_{\mathrm{tg}}$), bottom gate voltage ($V_{\mathrm{bg}}$) and the electrostatic potential drops $V_0$, $V_1$, $V_2$. $e$ is the elementary charge. (c) Resistance of MLG measured at 200 K. This 2D color map can be divided into nine regions based on the trace of CNP and its inflection points (indicated by the dashed white lines) of MLG. The line cuts of MLG resistance are shown on the right panel of (c). (d) The band alignments of MLG and $Nb_3Cl_8$ across the nine regions shown in (c).

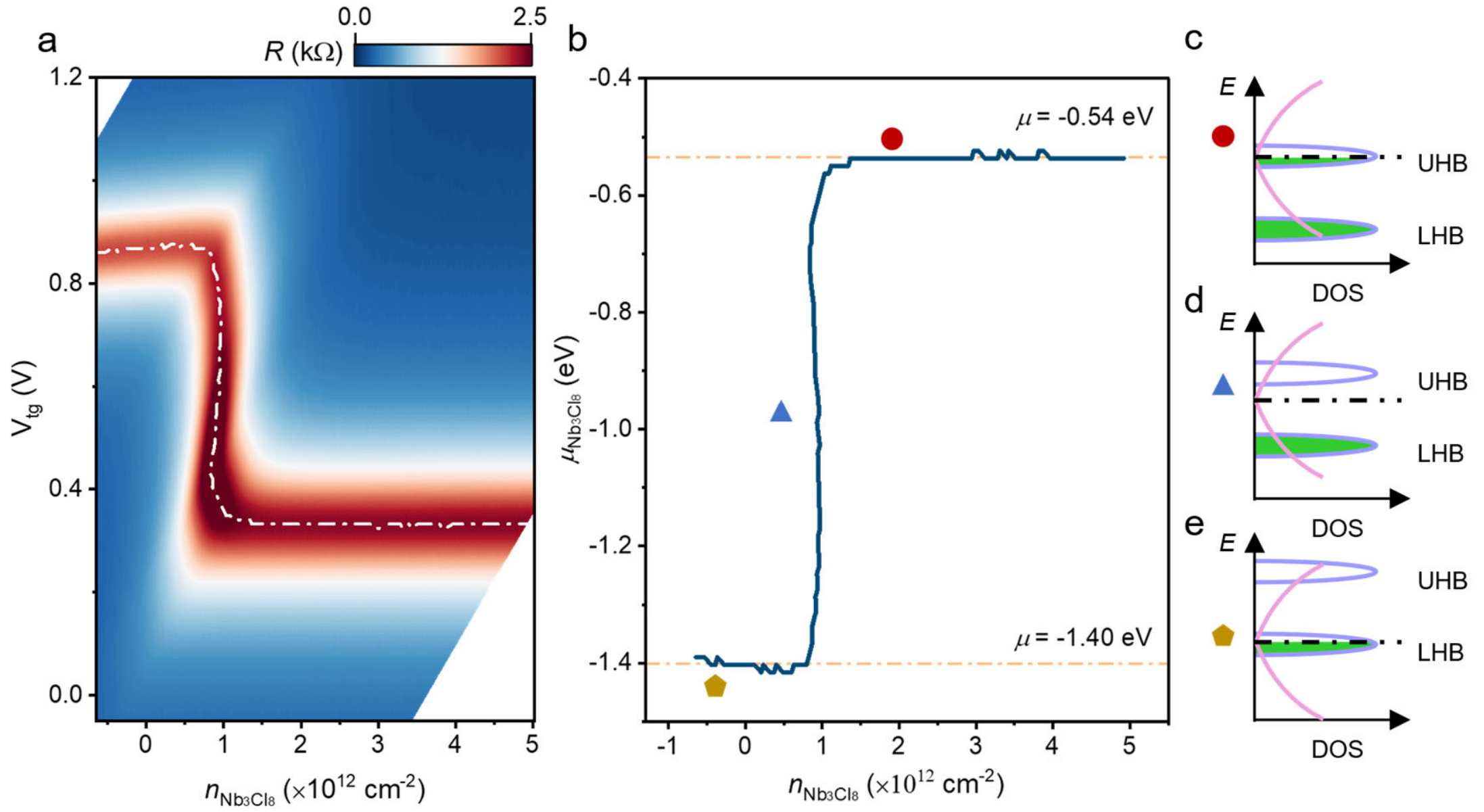

**Figure 3. The chemical potential of $Nb_3Cl_8$.** (a) The resistance of MLG versus $V_{\mathrm{tg}}$ and the carrier density of $Nb_3Cl_8$ at 200 K. The white lines indicate the CNP positions of graphene. By tracking the position of CNP, where all charge carriers reside in $Nb_3Cl_8$ layers, the relative chemical potential of $Nb_3Cl_8$ (linear with $V_{\mathrm{tg}}$) as a function of $n_{\mathrm{Nb_3Cl_8}}$ can be obtained. (b) The chemical potential of $Nb_3Cl_8$ versus its carrier density, extracted from the map in (a). The chemical potential of $Nb_3Cl_8$ is pinned at $-1.40$ eV on the hole side and $-0.54$ eV on the electron side, yielding a gap with its value equals to $\Delta\mu_{\mathrm{Nb_3Cl_8}} \sim 0.86$ eV. (c-e) The Hubbard band filling of $Nb_3Cl_8$. The Fermi level remains at the lower Hubbard band (e) when the carrier density is below $1 \times 10^{12}\ \mathrm{cm}^{-2}$, as the carrier density increases, the Fermi level rises to the Mott gap (d) and eventually reaches the upper Hubbard band (c).

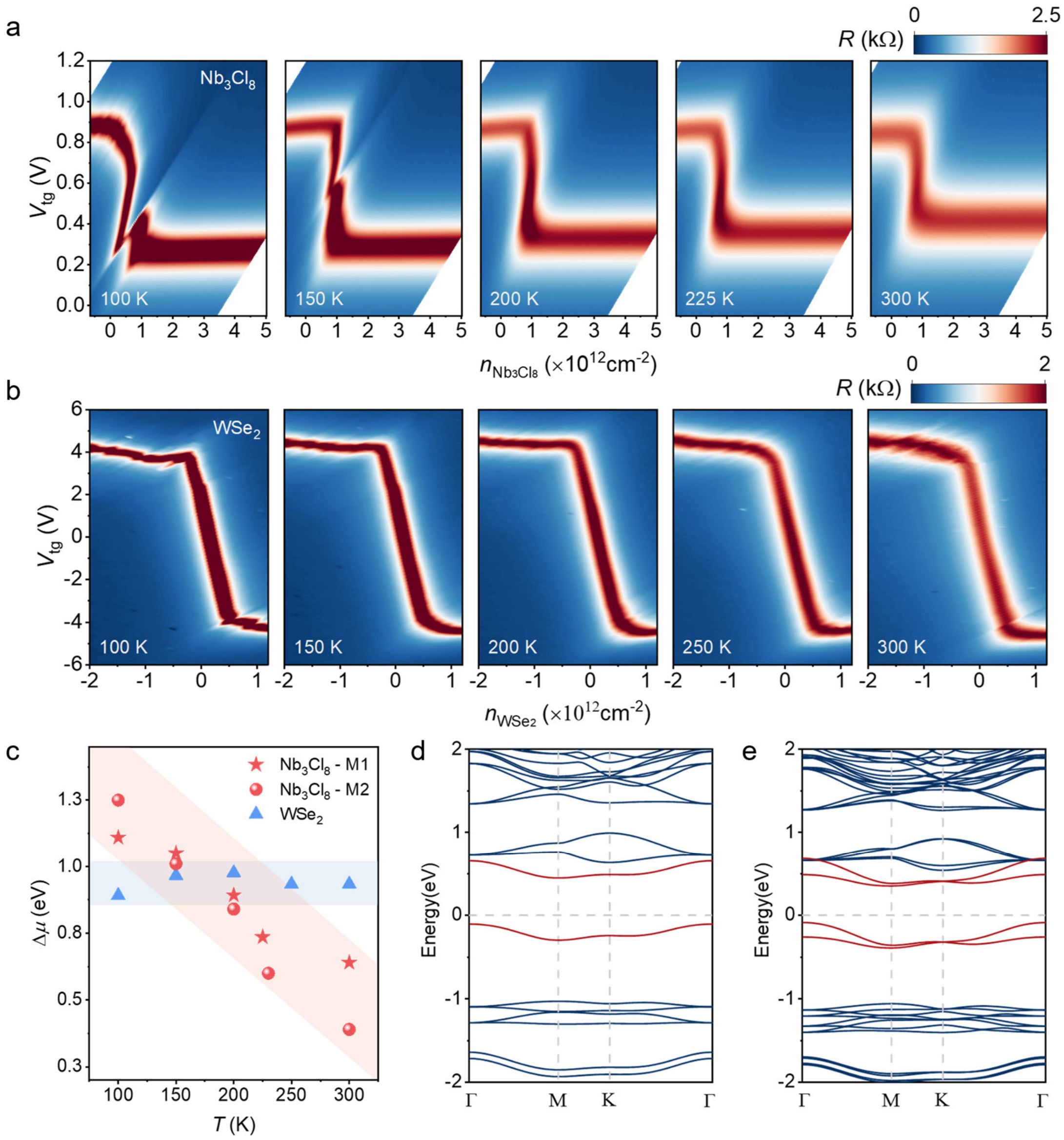

**Figure 4. The Mott insulating behavior of $Nb_3Cl_8$.** (a, b) Four-terminal resistance of MLG from measured from 100 K to 300 K on device MLG/hBN/$Nb_3Cl_8$ (a) and MLG/hBN/$WSe_2$ (b) heterostructures, respectively. The chemical potential of $Nb_3Cl_8$/bilayer $WSe_2$ can be obtained according to the CNP of MLG. (c) The gap of $Nb_3Cl_8$ and bilayer $WSe_2$ from 100 K to 300 K, calculated based on the difference in chemical potential between electron or hole doped regime. As temperature increases, the gap of $Nb_3Cl_8$ (represented by red stars and dots for devices M1 and M2, respectively) decreases sharply, in strongly contrast to that of $WSe_2$ (blue triangles), which is nearly independent on temperature. (d, e) The band structure of monolayer (d) and bulk (e) $Nb_3Cl_8$ after considering the electron correlation effect with Hubbard term $U_{eff} = 2$ eV. The strong electron correlations split the half-filled band near the Fermi level (shown in Figs. 1(c) and S4(a)) into upper and lower Hubbard bands, giving rise to a Mott gap.